\documentclass[pra,twocolumn,superscriptaddress,amsmath,showpacs,nofootinbib]{revtex4}

\usepackage{amsmath}
\usepackage{graphicx}
\usepackage{amssymb}
\usepackage{epsfig}
\usepackage{amsfonts}
\usepackage{color}

\def\fig#1{{#1}}

\newcommand{\Wuniv}{{W^{(4)}_{\rm univ}}}

\newcommand{\<}{\langle} \def\>{\rangle}
 
\newcommand{\tr}{\mathrm{tr}}
\def\ket#1{{\,|#1\rangle}}
\def\bra#1{{\langle#1|\,}}

\begin{document}

\title{Method for universal detection of two-photon polarization
entanglement}

\author{Karol Bartkiewicz}
\email{bark@amu.edu.pl} \affiliation{Faculty of Physics, Adam
Mickiewicz University, 61-614 Pozna\'n, Poland}
\affiliation{RCPTM, Joint Laboratory of Optics of Palack\'y
University and Institute of Physics of Academy of Sciences of the
Czech Republic, 17. listopadu 12, 772 07 Olomouc, Czech Republic }

\author{Pawe{\l} Horodecki}
\affiliation{Faculty of Applied Physics and Mathematics, Gda\'nsk
University of Technology, 80-233 Gda\'nsk, Poland}
\affiliation{National Quantum Information Centre in Gda\'nsk,
81-824 Sopot, Poland}

\author{Karel Lemr}
\affiliation{RCPTM, Joint Laboratory of Optics of Palack\'y
University and Institute of Physics of Academy of Sciences of the
Czech Republic, 17. listopadu 12, 772 07 Olomouc, Czech Republic }

\author{Adam Miranowicz}
\affiliation{Faculty of Physics, Adam Mickiewicz University,
61-614 Pozna\'n, Poland}

\author{Karol \.{Z}yczkowski}
\affiliation{Institute of Physics, Jagiellonian University, Ulica
Reymonta 4, 30-059 Krak{\'o}w, Poland} \affiliation{Center for
Theoretical Physics, Polish Academy of Sciences, Aleja
Lotnik{\'o}w 32/46, 02-668 Warsaw, Poland}

\begin{abstract}
Detecting and quantifying quantum entanglement of a given unknown
state poses problems that are fundamentally important for quantum
information processing. Surprisingly, no direct (i.e., without
quantum tomography) universal experimental implementation of a
necessary and sufficient test of entanglement has been designed
even for a general two-qubit state. Here we propose an
experimental method for detecting a collective universal witness,
which is a necessary and sufficient test of two-photon
polarization entanglement. It allows us to detect entanglement for
any two-qubit mixed state and to establish tight upper and lower
bounds on its amount. A different element of this method is the
sequential character of its main components, which allows us to
obtain relatively complicated information about quantum
correlations with the help of simple linear-optical elements. As
such, this proposal realizes a universal two-qubit entanglement
test within the present state of the art of quantum optics. We
show the optimality of our setup with respect to the minimal
number of measured quantities.
\end{abstract}



\date{\today}

\maketitle

\section{Introduction}

Quantum entanglement~\cite{Schr,EPR} is a fascinating phenomenon
considered to be one of the main resources in quantum information
and quantum engineering (for reviews, see Refs.~\cite{BZ,RMP,GT}).
In general, detecting entanglement in various physical scenarios
poses a significant problem. Most widely used methods are based on
measuring entanglement witnesses (see Ref.~\cite{RMP}), which are
efficient but, typically, not universal and require some
information about the state prior to its measurement. On the other
hand, by performing standard methods of quantum tomography of a
given state one obtains complete information about that state.
Thus, information concerning its entanglement can be extracted by
an explicit calculation, through the postprocessing of the
complete experimental data. However, full tomography requires
measuring a large number of parameters; this number scales with
the square of the total dimension of a measured state. Moreover,
there remains one conceptually fundamental question, namely what
is the minimal number of parameters that are experimentally
feasible (in the sense of, e.g., linear optics) that will
nevertheless provide complete information about quantum
entanglement independently of a general input state. This can be
viewed as a question about a quantum processor with a quantum
input (state) and a classical output (giving a yes or no answer or
some quantitative information about entanglement) with minimal
processing of classical (incoherent) information inside.

Early proposals regarding the detection and quantification of
quantum entanglement without state reconstruction were based on
the identification of polynomial moments. These methods made it
possible to retrieve information on entanglement from the data
spectrum of the partial transpose of the two- qubit Wootters
concurrence (see Refs.~\cite{PHEkert,PH,Carteret} for a
significant quantum-noise reduction). They enabled sharp two-qubit
entanglement tests, but required nonlinear postprocessing of the
data to retrieve the original information about entanglement.

Independently of the above-mentioned line of research, the concept
of collective entanglement witnesses~\cite{Horodecki03} made it
possible to construct collective observables for describing
entanglement quantitatively in experiments~\cite{Bovino}.
Moreover, the analysis of the concurrence of Ref.~\cite{Mintert}
(see also Ref.~\cite{ZHChen} for recent developments) eventually
led to a quantitative experimental estimation of entanglement in
terms of specific two-copy collective
witnesses~\cite{Aolita,Walborn}.

Another interesting example of collective entanglement witnesses
is a two-copy witness based on the geometric intuition of the
concept of metric~\cite{Badziag}. A number of multipartite tests
based on the nonlinear functions of simple multicopy observables
were developed~\cite{Huber10} and several other
quantitative~\cite{GRW07,EBA07,OH10,Sheng15}  and
qualitative~\cite{Badziag,Huber10,RHZ11,JMG,RPHZ12,RPHZ14} methods
of detecting entanglement without quantum tomography were
proposed. Nevertheless, these techniques, although quite powerful,
are not universal and the quality of their results depends on a
given state.

Experimental adaptive approaches~\cite{Park,Laskowski} were also
proposed for the case of two-qubits which we shall focus on in
this paper. Although these methods are an elegant improvement,
they do not satisfy the universality requirement.

Let us stress, however, that there exists a universal witness of
entanglement (UWE) for a two-qubit state, as introduced in
Ref.~\cite{Augusiak08} and defined here in Sec.~II. This UWE can
be measured by performing the joint measurements on the four
copies of a given state~\cite{Augusiak08}. However, so far no
experimental implementation for such a measurement has been
proposed. The aim of this paper is to propose a constructive
measurement procedure that outputs the mean value $\<\Wuniv\>$ of
the above witness for any two-qubit polarization state of a pair
of photons, thereby allowing us to detect the arbitrary quantum
entanglement of such systems. To our knowledge this is the first
experimental proposal of a universal (sharp) entanglement test
with (i) elementary (linear) optics and (ii) practically trivial
(direct substitution for polynomial) postprocessing of
experimental data. To be more specific, the procedure has the
unique advantage that it can be (probabilistically) utilized with
the help of just linear-optical methods involving only a sequence
of beam splitters and the Hong-Ou-Mandel (HOM) interference. Quite
remarkably, no polarizing beam splitters or phase rotators are
needed. This is especially important here because we consider only
the polarization-encoded qubits.

We start the presentation of our results with the analysis of the
properties of the UWE symmetries of the observables needed for
reproducing the three moments $\Pi_{i}$ ($i=2,3,4$). Then we shall
provide the optical HOM interference methods for reproducing the
values of the moments. Having found these values, one just needs
to substitute them into the polynomial~(\ref{Witness}) and to
check the sign of the final value.

This paper is organized as follows: In Sec.~II we recall the
definition of the UWE and show how the negativity and concurrence
are bounded by some functions of the UWE. In Sec.~III we present
the main idea how to measure the UWE. A detailed derivation of one
of our important formulas is given in the Appendix. In Sec.~IV we
relate our results to Makhlin's invariants and find the minimum
number of independent measurements required for detecting
entanglement. In Sec.~V we describe our proposal of an
experimental photonic implementation for the UWE detection. In
Sec.~VI, we present  two alternative implementations.  We discuss
our results and summarize in Sec.~VII.

\section{Universal entanglement witness and bounds on negativity and
concurrence}

\begin{figure}
\fig{\includegraphics[width=8cm]{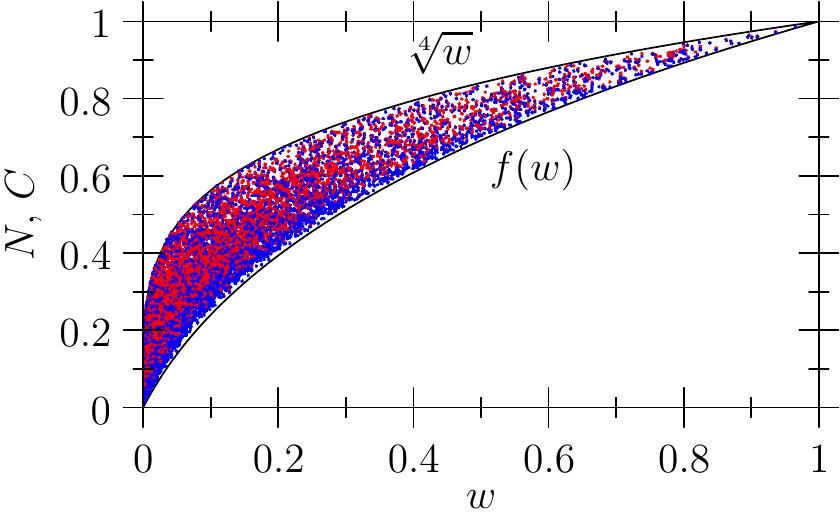}} \vspace*{-5mm}
\caption{\label{fig1} (Color online) Tight lower and upper bounds
for the two-qubit negativity $N$ and concurrence $C$ in terms of
the universal witness value $w=\max\big[0,-16\<\Wuniv\>\big]$. Red
(blue) dots correspond to the concurrence (negativity) for $10^4$
density matrices $\rho$ generated by a Monte Carlo simulation (see also \cite{Augusiak08}).}
\end{figure}

The UWE, provided in Ref.~\cite{Augusiak08}, is an operator
$\Wuniv$ such that its expected value corresponds to the
determinant of the partially transposed (marked by $\Gamma$)
two-qubit matrix $\rho$, i.e.,
\begin{eqnarray}
\det\rho^\Gamma & =&\<\Wuniv\>:=
\tr\big(\Wuniv\rho^{\otimes 4}\big) \nonumber \\
&=&\tfrac{1}{24}(1-6\Pi_4 + 8\Pi_3 + 3\Pi_2^2 - 6\Pi_2),
\label{Witness}
\end{eqnarray}
which is given in terms of the moments $\Pi_n =
\tr[(\rho^\Gamma)^n]$. For convenience, we refer to the observable
$\Wuniv$, but also to its expectation value $\<\Wuniv\>$ as the
UWE. It follows from the positive partial transpose (PPT)
criterion that a two-qubit state is entangled if and only if
$\<\Wuniv\> < 0$. The explicit form of this witness, which is the
mean value of the Hermitian observable $\Wuniv$ on the four copies
$\rho^{\otimes 4}$ of qubit pairs in a given state $\rho$, is
explicitly provided in Ref.~\cite{Augusiak08} and constructed from
permutation matrices. The main advantage of this UWE compared to
other universal methods of two-qubit entanglement detection is
that this is a linear observable that does not require solving
(unfolding) nonlinear polynomial equations, which are more
sensitive to errors, to obtain the information about a given state
(see Refs.~\cite{PHEkert,PH,Carteret}). Another advantage of this
witness is that its rescaled value
$w:=\max\big[0,-16\<\Wuniv\>\big]$ provides tight upper and lower
bounds~\cite{Augusiak08} on the negativity $N(\rho)$ and
concurrence $C(\rho)$ of a two-qubit state $\rho$,
\begin{equation}\label{eq:bounds}
f(w) \leq N \leq C \leq \sqrt[4]{w},
\end{equation}
where $f(w)$ is the inverse of the polynomial $w(C)=C(C+2)^{3}/27$
on the interval $C\in [0,1]$. We recall that the negativity~$N$ of
a two-qubit state $\rho$ can be defined as~\cite{RMP}:
\begin{equation}
N(\rho) = 2\max\{0,-\min[{\rm eig}(\rho^\Gamma)]\},
\end{equation}
i.e., via the minimum (negative) eigenvalue of the partially
transposed density matrix $\rho^\Gamma$, while the concurrence~$C$
of a two-qubit state $\rho$ is given by~\cite{Wootters98}:
\begin{equation}\label{eq:C}
C({\rho})=\max \Big(0,-\sum_j\lambda_j+2\max_j\lambda_j\Big),
\end{equation}
where $\{\lambda^2 _{j}\} = \mathrm{eig}[{\rho }({\sigma
}_{2}\otimes {\sigma }_{2}){\rho}^{\ast }({ \rho }_{2}\otimes
{\sigma }_{2})]$ and ${\sigma }_{2}$ is the Pauli operator. The
lower bound $f(w)$ can be given explicitly in terms of the
universal witness value $w$ as follows
\begin{eqnarray}
  f(w) = \tfrac{1}{2}\left(-3 + \sqrt{z} + \sqrt{3 - z +
\tfrac{2}{\sqrt{z}}}\right), \label{N1}
\end{eqnarray}
where $z = 1 +x - 36 w/x$, and
\begin{eqnarray}
x = 3 \sqrt[3]{2 \sqrt{w^2 (16 w+1)}-2 w}.
\end{eqnarray}
This lower bound and the upper bound are shown in Fig.~\ref{fig1}.

\section{How to measure universal entanglement witness: Principle idea}

\begin{figure}
\fig{\includegraphics[width=8cm]{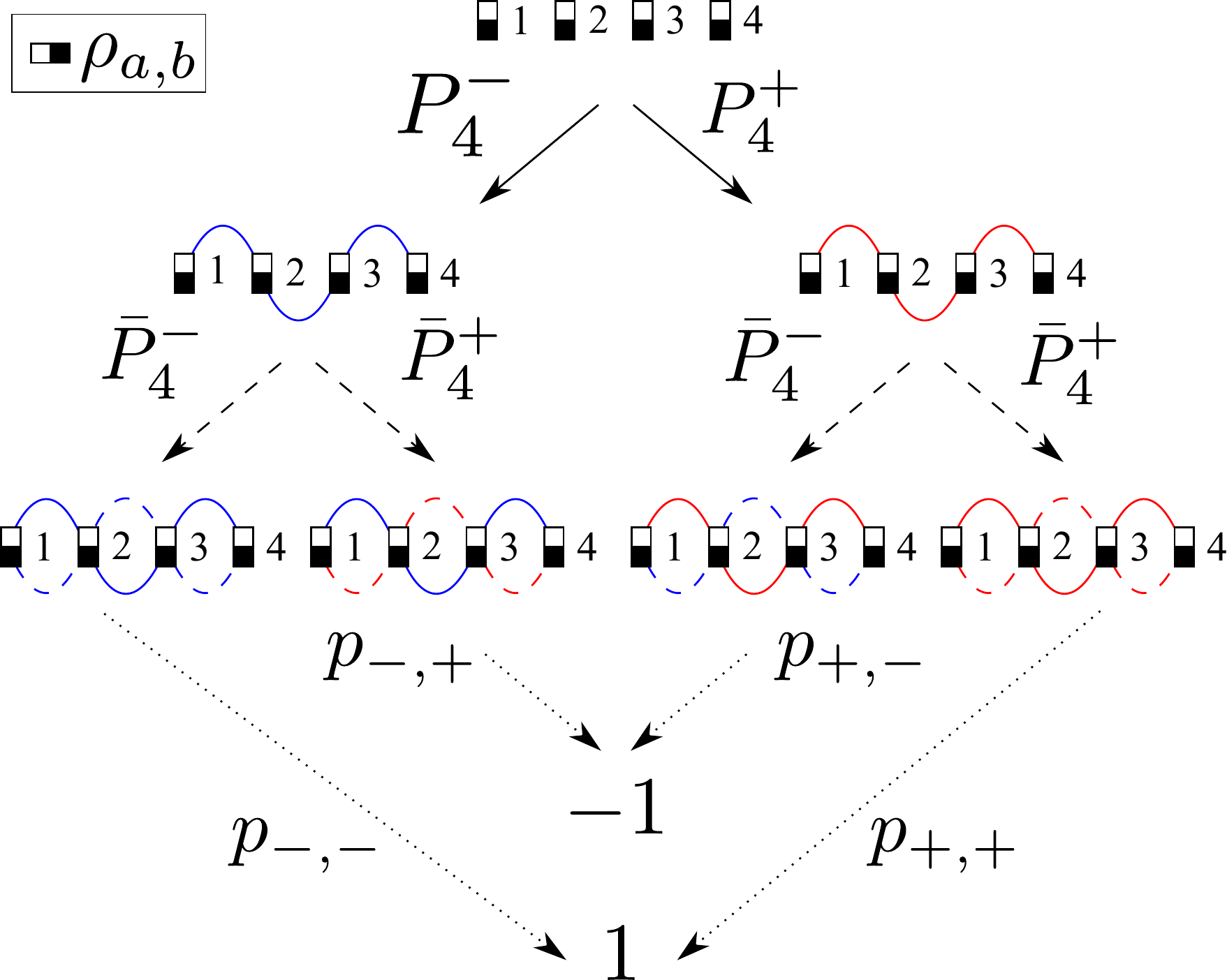}} \caption{\label{fig2}
(Color online) Conceptual diagram for measuring the moment
$\Pi_4$. The four copies of a given two-photon state $\rho_{a,b}$
are marked by small rectangles, where the white (black) part
corresponds to the $a$ ($b$) photon. Solid (dashed) curves connect
photons that are measured simultaneously at the first (second)
stage. The state $\rho_{a,b}$ is split into the symmetric and
antisymmetric parts by the projectors $P_4^+$ and $P_4^-$ (solid
arrows), respectively. This can be considered as a transformation
that is deterministic in principle. Next, the two branches are
split again into the symmetric (by the projector $\bar P_4^+$) and
antisymmetric ($\bar P_4^-$) parts (dashed arrows). These
projectors are applied to the two subgroups of qubits, as
indicated by the red (blue) curves for the symmetric
(antisymmetric) subspace projections. There are four possible
outcomes of this procedure. The events, indicated by dotted
arrows, correspond to measuring the value $+1$ ($-1$) with the
probability $p_{\pm,\pm}$ ($p_{\mp,\pm}$), where
$p_{+,+}+p_{+,-}+p_{-,+}+p_{-,-}=1$ and $\langle\Pi_n \rangle=
p_{+,+} - p_{+,-} - p_{-,+}+p_{-,-}$. The same procedure can be
used for measuring the lower-order moments $\Pi_n$ for $n=2,3$ if
the last $(4-n)$ copies of $\rho_{a,b}$ are removed. }
\end{figure}

In order to directly determine the witness $\<\Wuniv\>$, we can
measure all the moments $\Pi_n = \tr(\rho^\Gamma)^n$ separately.
We know how to measure $\Pi_2$ (see
Ref.~\cite{Bartkiewicz13fidelity}), since it is equivalent to the
purity of $\rho = \rho_{a_1,b_1}$. The remaining problem is how to
measure $\Pi_{3,4}$. As already mentioned, these moments were
originally reproduced as the mean values of the observables
constructed from permutation operators. However, the direct
measurement of the observables seems to be difficult due to their
relatively complicated structure. Fortunately, we can express the
moments $\Pi_{n}$ (for $n=3,4$) differently, i.e., by decomposing
the $n$th cycle into the products of inversions (the SWAP
operations $S$) as
\begin{eqnarray}
\Pi_{n} &=& \tr(A_{n}B_{n}\rho^{\otimes n}),
\end{eqnarray}
where $n=2,3,4$ and
\begin{eqnarray}
A_2 &=& S_{a_1,a_2}\otimes I_{b_1}\otimes I_{b_2},\nonumber \\
B_2 &=& I_{a_1}\otimes I_{a_2} \otimes S_{b_1,b_2},\nonumber \\
A_3 &=& S_{a_1,a_2}\otimes I_{a_3}\otimes I_{b_1}\otimes S_{b_2,b_3},\nonumber \\
B_3 &=& I_{a_1}\otimes S_{a_2,a_3}\otimes S_{b_1,b_2}\otimes I_{b_3},\nonumber \\
A_4 &=& S_{a_1,a_2}\otimes S_{a_3,a_4}\otimes I_{b_1}\otimes S_{b_2,b_3}\otimes I_{b_4},\nonumber \\
B_4 &=& I_{a_1}\otimes S_{a_2,a_3}\otimes I_{a_4}\otimes
S_{b_1,b_2}\otimes S_{b_3,b_4},
\end{eqnarray}
together with the swap operator
$S=\ket{HH}\bra{HH}+\ket{HV}\bra{VH}+\ket{VH}\bra{HV}+\ket{VV}\bra{VV}$
and the single-qubit identity operator $I$,  where $\ket{H}$ and
$\ket{V}$ are the states of the horizontally and vertically
polarized photons, respectively. The products $A_nB_n$ are not
Hermitian for $n=3,4$, so they cannot be measured directly.
However, the operators $A_n$ and $B_n$ taken separately are
Hermitian and have other useful properties, i.e.,
$\tr(A_nB_n\rho^{\otimes n}) =\tr(B_nA_n\rho^{\otimes n})$ and
$A_n^2 =B_n^2=I^{\otimes{n}}$. By applying these properties we can
express higher-order moments of the partially transposed matrix as
\begin{equation}
\Pi_{n} = \tfrac{1}{2}\tr[(A_{n} + B_{n})^2\rho^{\otimes
n}] -1, \ \ n=3,4.
\end{equation}
Note that this method displays some analogy to the method of
calculating the collective spin of two parties. Let us define
$X_n=(A_n + B_n)^2$. Then, in order to measure these two moments,
we have to perform projections on the eigenspaces of $X_3$ and
$X_4$. The implementation of these operations might be difficult
for two reasons: (i) the large number of different eigenvalues of
the operators $X_n$ (positive operator valued measures), and (ii)
the complicated structure of the corresponding eigenspaces with
the eigenvectors corresponding to entangled multiqubit states.
Fortunately, the operator $X_3$ has only two different eigenvalues
$(1,4)$, resulting in two eigenspaces; and the operator $X_4$ has
only three different eigenvalues $(0,2,4)$, hence it has three
eigenspaces. Therefore, one has to perform only the measurement of
five projections on some of the eigenspaces to measure both
$\Pi_3$ and $\Pi_4$. The remaining problem is to find the
eigenspaces and associate them with the specific settings of a
multiphoton interferometer. Measuring the second moment $\Pi_2$
requires two projections. Thus, the complete measurement of
$\Wuniv$ can be decomposed into seven projections (this value may
be even lower if some optimization is applied) onto subspaces
spanned by highly entangled multiqubit states, which is twice as
efficient as a full two-qubit tomography. There is, however,
another method of measuring the products of $A_n$ and $B_n$ for
$n=3,4$ that is better regarding the complexity of these
projections. We can express $A_n=P_n^+ - P_n^-$ in terms of the
projectors onto the symmetric ($P_n^+$) and antisymmetric
($P_n^-$) subspaces. Then, as shown in the Appendix, we have
\begin{equation}
\Pi_n = \tr[B_n P_n^+\rho^{(n)} P_n^+] - \tr[B_n P_n^-\rho^{(n)}
P_n^-] \label{eq6}
\end{equation}
for an arbitrary $\rho$. It is convenient to define $P^\pm_{m,n}=
\tfrac{1}{2} (I_{a_m}\otimes I_{a_n} \pm S_{a_m,a_n})$ and $\bar
P^\pm_{m,n}= \tfrac{1}{2} (I_{b_m}\otimes I_{b_n} \pm
S_{b_m,b_n})$. Then the symmetric ($P_n^+$) and antisymmetric
($P_n^-$) projectors for $n=3,4$ read as
\begin{eqnarray}
P_3^\pm &=&  P^\mp_{1,2}\otimes \bar P^-_{2,3} +
P^\pm_{1,2}\otimes \bar P^+_{2,3},\\
P_4^\pm &=& P^{\mp}_3\otimes P^-_{3,4}+P^{\pm}_3\otimes
P^+_{3,4}.
\end{eqnarray}
For the operator $B_n= \bar P_n^+ -  \bar P_n^- $, we can apply the
same procedure but with the subsystems of the multiqubit density
matrix swapped as $a\leftrightarrow b$. Then, we have
\begin{eqnarray}
\Pi_n =\tr  \bigl(\bar P_n^+ Q\bar P_n^+\bigr) -
\tr\bigl(\bar P_n^-Q\bar P_n^-\bigr),
\end{eqnarray}
where $Q =  P_n^+\rho^{(n)} P_n^+ -P_n^-\rho^{(n)} P_n^- $, which
means that
\begin{eqnarray}
\Pi_n &=& \sum_{x,y=0}^1 (-1)^{x+y} \tr[  \bar P_n^x  P_n^y
\rho^{(n)} P_n^y  \bar P_n^x ], \label{eq10}
\end{eqnarray}
where $\bar P_n^{0} = \bar P_n^{+}$ and $\bar P_n^{1} = \bar
P_n^{-}$. Thus, it appears to be more convenient to project
$\rho^{(n)}$ onto the symmetric or antisymmetric subspace of $A_n$
first and then to measure $B_n = \bar P_n^{+} -\bar P_n^{-}$, as
shown in Fig.~\ref{fig2}.

\begin{figure}
\fig{\includegraphics[width=8.5cm]{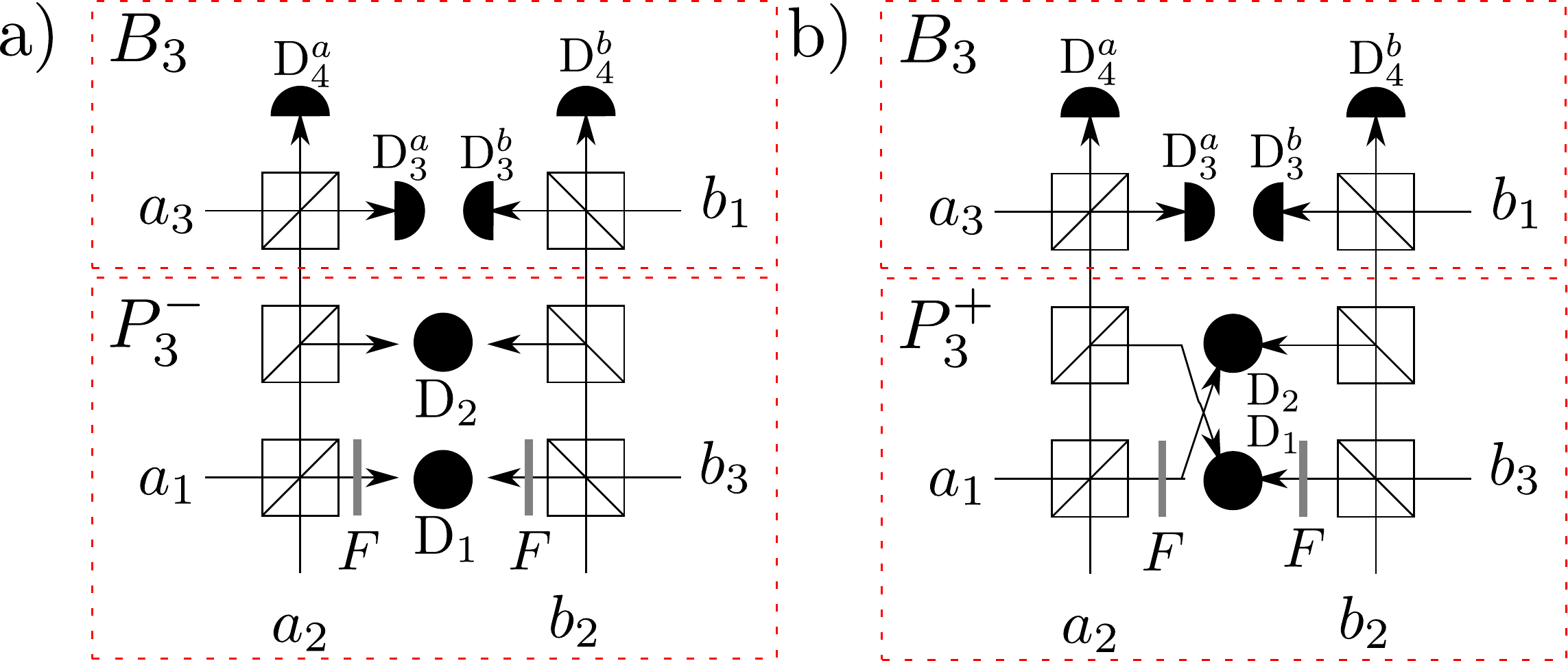}}
\caption{\label{fig3} (Color online) Linear-optical setup for
measuring $B_3P^\pm_3$ directly. The principle of its operation is
based on the fact that a beam splitter (BS) performs the
projection $P^+$ ($P^-$) if the incoming photons in $b_1$ and
$b_2$ are bunched (antibunched). If they are bunched, at least one
photon is passed to the detection mode of the first BS. Next, if
no photon is passed to the detection mode of the second BS, the
output state is $\tr_{b_1} (P^-\rho_{b_1,b_2}P^-)$. However, if
there is no photon in the first detection mode and there is a
photon in the second detection mode, the output state becomes
$\tr_{b_1}(P^+\rho_{b_1,b_2}P^+)$. The setup works if both the
detectors $D_1$ and $D_2$ register a photon, and it is unknown
from where the photons have arrived. The $B_3$ part is implemented
by distinguishing between $P^+$ and $P^-$ by means of detecting
bunching and antibunching, respectively. To guarantee the
optimality of this setup with respect to the minimal number of
measured quantities, the information about parities of individual
photon pairs should be erased as described in Sec. VI.B. Here we
assume that this is done by using photon-number-resolving
detectors $D^{a(b)}_4$, but this can also be done
probabilistically using bucket
detectors~\cite{Bartkiewicz13fidelity, Bartkiewicz13discord}.
Thus, assuming perfect detectors, one needs two measurements to
determine $\Pi_3$. Finally, neutral density filters $F$ of the
transmittance 1/2 ensure that the setup works with probability
1/16. }
\end{figure}

\begin{figure}
\fig{\includegraphics[width=7.4cm]{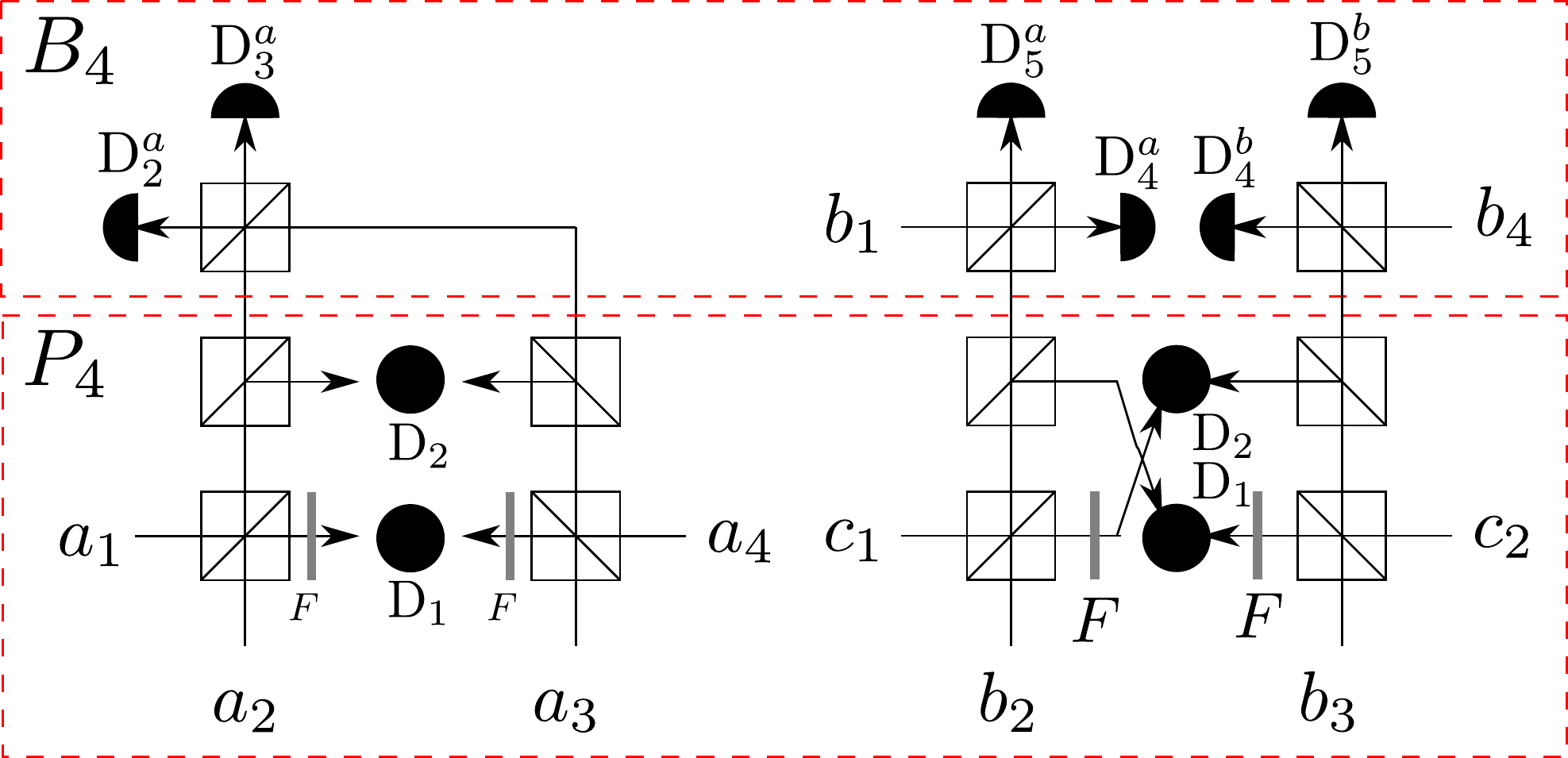}}
\caption{\label{fig4} (Color online) Linear-optical setup for
direct measuring $B_4P^\pm_4$. The principle of its operation is
similar to that in Fig.~\ref{fig3}. Here, for simplicity, we
assume that the detectors $D_1$ and $D_2$ can distinguish between
the even and odd numbers of photons. Note that this assumption is
irrelevant in the setups discussed in Sec.~VI. If the even (odd)
number of photons is passed to detector $D_1$, the measurement
result is that of the $B_4$ measurement multiplied by $1$ ($-1$).
The ancillary modes $c_1$ and $c_2$ are prepared in the
polarization singlet state. Note that the right-hand-side module
corresponds to that shown in Fig.~\ref{fig3}(b) but with the
replaced notation for the input modes and detectors:
$(a_1,a_2,a_3,b_1,b_2,b_3) \rightarrow (c_1,b_2,b_1,b_4,b_3,c_2),$
and $(D^a_3,D^b_3,D^a_4,D^b_4) \rightarrow
(D^a_4,D^b_4,D^a_5,D^b_5)$, respectively.}
\end{figure}

\section{Makhlin's invariants and minimal number of
independent measurements}

The moments $\Pi_{\lbrace  1,2,3,4 \rbrace}$ are interrelated. In
order to demonstrate this property, let us express the two-qubit
density matrix $\rho$ in terms of the Pauli matrices $\sigma_i$
for $i=1,\,2,\,3$ and the single-qubit identity matrix $\sigma_0$.
The resulting matrix reads
\begin{equation}
\rho = \tfrac{1}{4}\sigma_0\otimes\sigma_0 +
\tfrac{1}{2}s_i\sigma_i\otimes\sigma_0 +
\tfrac{1}{2}p_j\sigma_0\otimes\sigma_j +
\beta_{ij}\sigma_i\otimes\sigma_j,
\end{equation}
where the  elements of the correlation matrix $\hat{\beta}$ are
$\beta_{ij} = \mathrm{tr}[(\sigma_i \otimes\sigma_j)\rho]$ and the
Bloch vectors $\mathbf{s}$ and $\mathbf{p}$ have the following
elements of $s_i = \mathrm{tr}[({\sigma}_i\otimes{\sigma}_0)\rho]$
and $p_j = \mathrm{tr}[(\sigma_0\otimes{\sigma}_j)\rho]$,
respectively. It can be directly shown, after tedious
calculations, that
\begin{eqnarray}
\Pi_1 &=& 1,\nonumber \\
4\Pi_2 &=&   1 + x_1,\nonumber\\
16\Pi_3 &=& 1 + 3x_1 + 6x_2,\nonumber\\
64\Pi_4 &=& 1 + 6x_1+  24x_2 + x_1^2 + 2x_3 +4x_4,
\end{eqnarray}
where
\begin{eqnarray}
x_1 &=&  I_2 + I_4 + I_7,\quad
x_2 = I_1 + I_{12},\nonumber\\
x_3 &=& I_2^2 - I_3, \quad\quad\quad x_4 = I_5 + I_8 + I_{14} +
I_4I_7,
\end{eqnarray}
are functions of nine local invariants of the two-qubit matrix
$\rho$ as defined by Makhlin in Ref.~\cite{Makhlin00}, i.e.,
$I_1=\det\hat\beta$, $I_2=\mathrm{tr}(\hat\beta^T\hat\beta)$,
$I_3=\mathrm{tr}(\hat\beta^T\hat\beta)^2$, $I_4=\mathbf{s}^2$,
$I_5=[\mathbf{s}\hat\beta]^2$, $I_7 = \mathbf{p}^2$, $I_8 =
[\hat\beta\mathbf{p}]^2$, $I_{12} =
\mathbf{s}\hat\beta\mathbf{p}$, and $I_{14} =
e_{ijk}e_{lmn}s_ip_l\beta_{jm}\beta_{kn}$, where $e_{ijk}$ is the
Levi-Civit\`a symbol. It is apparent that only six (instead of
nine) linear combinations of Makhlin's invariants need to be
measured to estimate the values of $x_n$ for $n=1,..4$. These
invariants read
\begin{eqnarray}
y_1  &=& I_2,\quad y_2 = I_3,\quad y_3  = I_4, \quad y_4  =  I_7,
\quad
\nonumber \\
y_5  &=& I_1 + I_{12,},\quad y_6  =  I_5 + I_8 + I_{14}.
\end{eqnarray}
Thus, in order to detect the entanglement via $\det\rho^\Gamma$,
one needs to measure exactly six instead of nine independent
linear combinations of fundamental invariants. It also happens
that this is also the minimal number of independent fundamental
quantities describing the negativity of an arbitrary two-qubit
state~\cite{Bartkiewicz15}.

\section{A proposal of experimental implementation}

The analyzed projections $P_n^\pm$   are not the products of
projections (except $n=2$),  thus they cannot be implemented by
local (two-qubit) operations.  Let us note that the $P_n^\pm$
projectors split a collective multiqubit state into the states of
positive and negative parities. This technique was also applied
to, e.g., the cluster-state preparation~\cite{Barrett05}.

We can, however, measure $B_3P_3^\pm$ directly as shown in
Fig.~\ref{fig3}. Note that $B_3$ can be measured using only beam
splitters and photon detectors analogously to the methods applied
in Refs.~\cite{Bartkiewicz13fidelity,
Bartkiewicz13chsh,Bartkiewicz13discord}. In Figs.~\ref{fig3}
and~\ref{fig4} we show a simple implementation of a $B_n$ block
(for $n=3,\,4$). To measure the three parameters from
Fig.~\ref{fig2} (four parameters without normalization) instead of
using the $B_n$ block we can reuse the $P^\pm_n$ part of the
relevant setup to perform a $\bar{P}^\pm_n$ projection (see
Sec.~VI).

\begin{figure}
\includegraphics[scale=0.5]{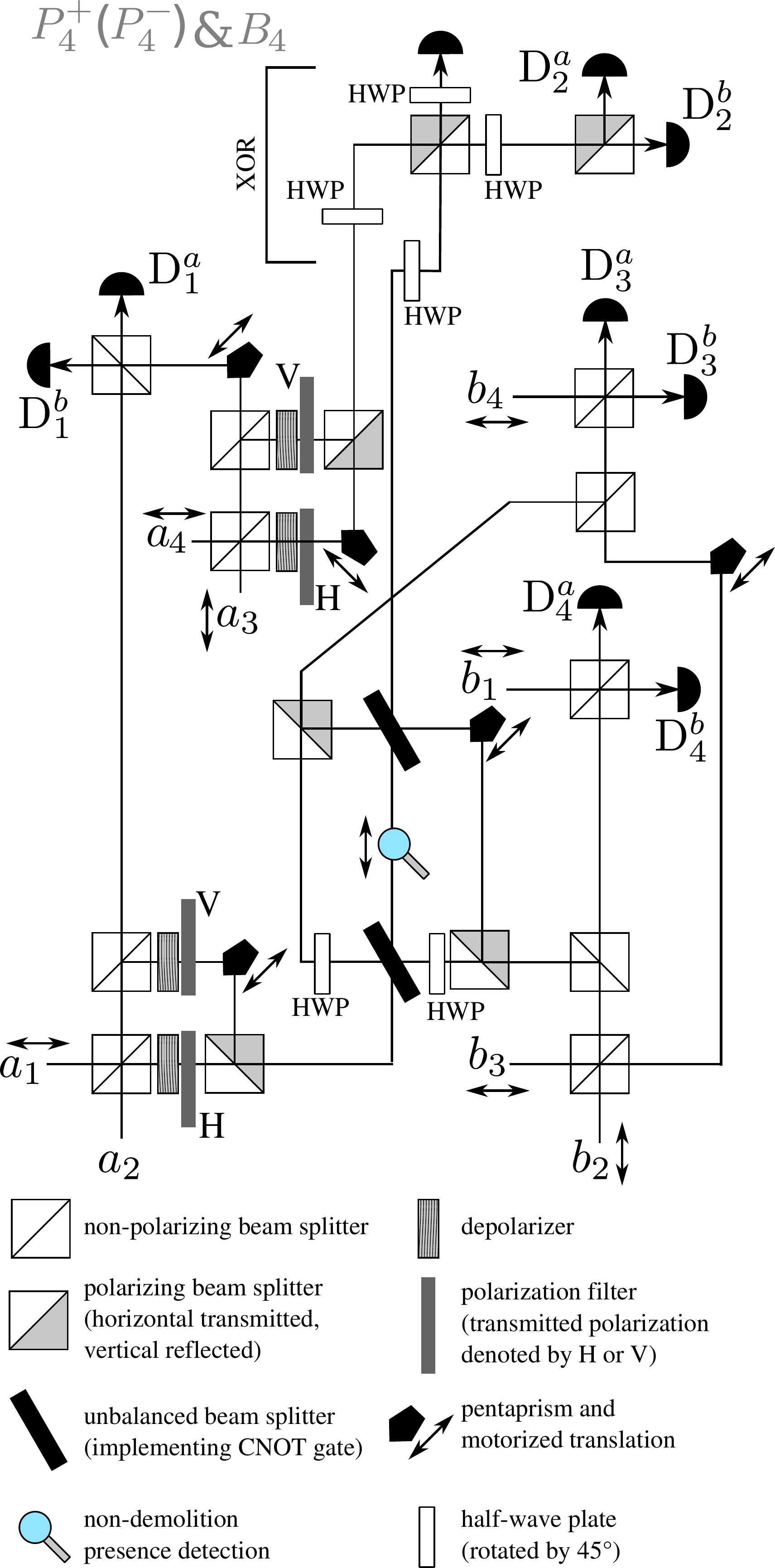}

\caption{\label{fig5}  (Color online) Alternative setup for the
direct measurement of $B_4P_4^\pm$. The method uses three known
quantum gates: the XOR, the CNOT (i.e., the reversible XOR), and
the nondemolition presence detection gate. All these gates can be
implemented using linear optics only. The XOR gate can easily be
constructed using a polarizing beam splitter and a set of
half-wave plates (HWP) (see, e.g., Ref.~\cite{Pittman05}). The
CNOT gate can be built using a special partially polarizing beam
splitter (see, e.g., Ref.~\cite{Kiesel05}). Finally, the
nondemolition presence detection can be achieved with the
assistance of two ancillae~\cite{Bula13QND} in the polarization
singlet state. The method is successful if there are two photons
registered by each pair of the photon-number resolving detectors
$D_1$, $D_3$, and $D_4$, while one photon is detected by either
$D_2^a$ or $D_2^b$. It is easy to show that the photon detection
by $D_2^a$ corresponds to the $B_4P_4^-$ measurement, while the
detector $D_2^b$ heralds the $B_4P_4^+$ measurement.}
\end{figure}

For any $P_n^\pm$, two qubits $a_1$ and $a_n$ ($b_n$) for even
(odd) $n$ can be destroyed in this process. In the most complex
case of $P_4^\pm$, we have to perform a nondestructive parity test
on six qubits, where two of them can be destroyed before measuring
$B_4$. The measurement of $B_4P_4^\pm$ is more challenging than
that of $\Pi_3$. In the comparison to the $\Pi_3$ setup, the main
difficulty here is the necessity to condition the outcome of
nondestructive measurements on $b_2$ and $b_3$. This is because
both $b_2$ and $b_3$ are required in the latter part of the
$\Pi_4$ measurement. We can solve this problem by using ancillary
photons prepared in the polarization singlet state in the modes
$c_1$ and $c_2$.  The corresponding setup is shown in
Fig.~\ref{fig4}. In some experimental approaches, this setup can
be further simplified by applying, e.g., the time-bin
methods~\cite{Marcikic02,Bartkiewicz13discord}. Our alternative
proposal is discussed in the next section and shown in
Fig.~\ref{fig5}. Finally, note that the relevant moments $\Pi_n$
can be measured with the subblocks of the setup in Fig.~\ref{fig4}
if some information is ignored.

\section{Alternative implementations}

\subsection{Alternative setup for measuring $\Pi_4$}

The setup depicted in Fig.~\ref{fig4} is fairly efficient and
experimentally not overly demanding. Unfortunately, it requires
the photon-number parity measurement, which (i) is experimentally
challenging and (ii) has photon losses in the setup that can
result in incorrect measurement outcome. Especially the second
limitation hinders the implementation of the method since photon
losses and imperfect detection efficiencies are unavoidable in
real experiments.

To overcome this problem, in addition to the time-bin approach as
already mentioned, we have devised another setup, as shown in
Fig.~\ref{fig5}, for the direct measurement of the $B_4P_4^\pm$
term. In contrast to the previous setup, shown in Fig.~\ref{fig4},
no parity measurement is required. On the other hand, the new
setup is much more complex and requires interferometric stability
(at some places). The idea behind the method is to use two quantum
gates: the controlled-NOT (CNOT)
gate~\cite{Eisert00,Pittman03,Kiesel05,Lemr_PRL106} and the
exclusive OR (XOR) gate~\cite{Pittman05}. In order to join two
CNOT gates, it is also required to introduce the nondemolition
photon presence detection gate, which uses two additional ancillae
in a Bell state~\cite{Kok02,Bula13QND}. By heralding the presence
of a qubit, this gate informs that the preceding CNOT operation
was successful. The entire measurement method is successful if two
photons are detected by each detector pair among $D_1$, $D_3$ and
$D_4$, while one photon is detected by either of the $D_2$
detectors and the presence detection gate also heralds a photon.
Further, if the detector $D_2^a$ fires, the method performs the
$B_4P_4^-$ measurement, while if the photon impinges on the
$D_2^b$ detector, the setup implements the $B_4P_4^+$ measurement.

\subsection{Alternative implementation of the $B_{3(4)}$ block}

If we take a look at the $B_{n}$ blocks with $n=3,4$ in
Figs.~\ref{fig3} and~\ref{fig4}, respectively, we will discover
that we check for bunching and antibunching separately for each
photon pair. In addition to the information about the outcome of
the $\bar{P}^{\pm}_{n}$ projections, we obtain the information
about which of the photon pairs is bunched or antibunched. We do
not use the which-pair information in any way (we just need to
know how may  pairs bunched or antibunched) and this measurement
is not difficult to implement. However, one may argue that we gain
more knowledge from our measurement that is necessary to measure
$\Pi_{n}$.

To perform the $B_{n}$ measurement and not to distinguish between
the pairs of photons one would have to use the same block as for
the ${P}^{\pm}_{n}$ measurement and to swap the modes
$a\leftrightarrow b$ to perform the $\bar{P}^{\pm}_{n}$
projections. This procedure would result in the four separate
detection events: ${P}^{+}_{n}\bar{P}^{+}_{n}$,
${P}^{+}_{n}\bar{P}^{-}_{n}$, ${P}^{-}_{n}\bar{P}^{+}_{n}$, and
${P}^{-}_{n}\bar{P}^{-}_{n}$  [see Eq.~(\ref{eq10})] associated
with the single observable $\Pi_{n}$. This number of the detection
events is now smaller that in the case of analyzing bunching and
antibunching for each pair at the original $B_{n}$ blocks. The
drawback of this method  is that it is more experimentally
challenging. However, by using the  time-bin approach, analogously
to that applied in Ref.~\cite{Bartkiewicz13discord}, we could
reuse the same physical $P^{\pm}_{n}$ setup to measure
$\bar{P}^{\pm}_{n}$ at a later moment of time. To summarize, we
may iterate the block measuring $P^{\pm}_{n}$ to get exactly the
statistics corresponding only to four exclusive events.

\section{Discussion and conclusions}

We showed how to directly measure the universal entanglement
witness $\Wuniv$ by using four copies of an arbitrary two-qubit
state $\rho_{a,b}$ of a two-photon polarization state. Our
approach consists of three measurements associated with three
moments of the partially transposed matrix $\rho^\Gamma_{a,b}$,
i.e., $\Pi_{n}=\tr[(\rho^\Gamma)^n]$ for $n=2,3,4$. The key issue
is to calculate the number of parameters that were estimated in
the process of measurement. Figure~\ref{fig2}  shows us that we
can estimate two independent quantities for each of the three
moments $\Pi_{\{2,3,4\}}$, which after normalization become the
probabilities $p_+=p_{-,-} + p_{+,+}$ and $p_-=p_{-,+}+p_{+,-}$.
Consequently, the output of the setup provides six parameters that
are generally independent. Note that there are nine parameters if
the normalizations are included.

The moments $\Pi_{\lbrace  2,3,4 \rbrace}$ are interrelated. Each
higher moment is a function of lower moments and some additional
parameters. After tedious calculations, we showed that
$\Pi_{\lbrace 2,3,4 \rbrace}$ are functions of nine fundamental
Makhlin invariants~\cite{Makhlin00} of $\rho$.  The relevant
invariants are $I_{\lbrace 1,2,3,4,5,7,8,12,14\rbrace}$ (see
Sec.~IV). Under closer inspection we discovered that only six
fundamental quantities are needed to estimate the values of
$\Pi_{\lbrace  2,3,4 \rbrace}$, i.e., to detect entanglement via
$\det\rho^\Gamma$ (or to measure the negativity of an arbitrary
two-qubit state~\cite{Bartkiewicz15}). These are $y_1  = I_2$,
$y_2  = I_4$, $y_3  =  I_7$, $y_4  = I_1 + I_{12}$, $y_5  = I_5 +
I_8 + I_{14}$, $y_6 = I_3$.  Thus, our setup estimates no more
quantities than those.

Note that local qubit unitary operations have three relevant
independent real parameters (excluding global phase). Thus, the
number of parameters of $U_{A} \otimes U_{B}$ is six while the
total number of parameters of a mixed two-qubit state is 15. The
resulting $9=15-6$ parameters are exactly all the relevant ones
after introducing the $U_{A} \otimes U_{B}$ invariant equivalence
classes, and they correspond to the nine fundamental invariants.
This number of parameters can be further reduced by swapping the
subsystems of $\rho$.

This is probably the reason behind the minimalist character of
this method. Indeed, it requires no unitary operations, which may
reflect the symmetry of the problem under local unitary
operations. Because of its simplicity, we believe that the
presented setup paves the way for the first experimental
realization of a necessary and sufficient universal test of
entanglement.

Finally, let us underscore that the key feature responsible for
the success of the proposed approach is the sequential character
of measurements. It seems that this property of the setup reframes
the paradigm for entanglement, correlations, and any other
nonlocal (i.e., not depending solely on the reduced density
matrices of subsystems) property of quantum-state detection and/or
estimation in practice. As a result a more general problem can be
conceived of. Given only very specific measurement modules
(analogous to a beam splitter in the Hong-Ou-Mandel interference
experiment), which can be repeated in different subsystems, is it
possible to estimate nonlocal quantities and, if so, what setup
would minimize the number of required measurements?

We developed a general method of the measurement of
invariant-based moments of partially transposed density matrices.
Our detailed description of the method is focused on the detection
of the entanglement of two qubits. This method can be generalized
and applied to measure (at least some) moments of partially
transposed density matrices of higher-dimensional systems too.
Nevertheless, such detection setups can be, in general, more
complicated than the setups for quantum-state tomography. However,
let us stress again the main result of our paper, which is the
first proposal of an experimental entanglement detection without
performing a complete quantum-state tomography. Our method enables
us to detect the entanglement between two qubits in an arbitrary
state if and only if these qubits are entangled. The main idea of
our method is based on the measurement of the universal
entanglement witness, which is a necessary and sufficient
entanglement condition for an arbitrary state of only two qubits.
In this sense, our universal two-qubit entanglement detection is
not directly scalable for two-qudit or multiqubit systems.

\begin{acknowledgments}
We gratefully acknowledge the financial support of the Polish
National Science Centre through Grants No. DEC-2011/03/B/ST2/01903
(A.M.), DEC-2013/11/D/ST2/02638 (K.B.), and No.
DEC-2011/02/A/ST2/00305 (K.\.{Z}.), the Czech Science Foundation
(Grant No. 13-31000P) (K.L.), the Czech Ministry of Education
through Grant No. LO1305 (K.B.) and the Quasar: Era-Net Chist-Era
7FP UE (P.H.).
\end{acknowledgments}

\appendix

\section{Derivation of Eq.~(\ref{eq6})}

In order to derive Eq.~(\ref{eq6}), let us start by noting that
\begin{equation}
\tr(B_nA_n\rho^{\otimes n}) = \tr\left[B_nA_n(\rho^{\otimes
n})'\right],
\end{equation}
where the statistical operator
\begin{equation}
(\rho^{\otimes n})' = \tfrac{1}{2}(\rho^{\otimes n} +
A_n\rho^{\otimes n}A_n),
\end{equation}
can be implemented by alternating between the input state
$\rho^{\otimes n}$ and $A_n\rho^{\otimes n}A_n$ (qubits are
swapped according to the definition of $A_n$). The newly obtained
density matrix $(\rho^{\otimes n})' $ has a very important
property, i.e., it commutes with the operator $A_n$:
\begin{equation}
\left[A_n,(\rho^{\otimes n})' \right] =
\tfrac{1}{2}\left(\left[A_n,\rho^{\otimes n}\right]
+\left[\rho^{\otimes n},A_n\right]\right) = 0.
\end{equation}
Thus, the two operators have a common set of eigenvectors
$|\psi^{(n)}_m\>$ for $m=1,2,..,4^n$, i.e., they are both diagonal
in the same basis. Let us expand the expression
$\tr\left[B_nA_n(\rho^{\otimes n})'\right]$ using the diagonal
representations of the operators $A_n = \sum_k a^{(n)}_k |
\psi^{(n)}_k\>\< \psi^{(n)}_k|$, $B_n = \sum_l b^{(n)}_l  |
\phi_l\>\< \phi_l|$ and $(\rho^{\otimes n})' = \sum_m r^{(n)}_m  |
\psi^{(n)}_m\>\< \psi^{(n)}_m|$. By doing so, we arrive at
\begin{eqnarray}
&&\tr\left[B_nA_n(\rho^{\otimes n})'\right]\nonumber\\
&&=\sum_{k,l,m} \tr (a^{(n)}_k b^{(n)}_l r^{(n)}_m   | \phi^{(n)}_l\>\< \phi^{(n)}_l| \psi^{(n)}_k\>\< \psi^{(n)}_k|\psi^{(n)}_l\>\<\psi^{(n)}_l|)\nonumber{} \\
&&= \sum_{k,l} a^{(n)}_k b^{(n)}_l r^{(n)}_k   |\< \phi^{(n)}_l|
\psi^{(n)}_k\>|^2.
\end{eqnarray}
This is equivalent to a measurement strategy consisting of
measuring $A_n$  first, and then measuring $B_n$, which can be
expressed as
\begin{eqnarray}
&&\hspace{-1cm}\sum_k\tr\left[B_n  a^{(n)}_k |\psi^{(n)}_k\>\<\psi^{(n)}_k| (\rho^{\otimes n})' |\psi^{(n)}_k\>\<\psi^{(n)}_k|\right]\nonumber \\
&&= \sum_{k,l} a^{(n)}_k b^{(n)}_l    |\< \phi^{(n)}_l|\psi^{(n)}_k\>|^2\<\psi^{(n)}_k| (\rho^{\otimes n})' |\psi^{(n)}_k\>\nonumber\\
&&=\sum_{k,l} a^{(n)}_k b^{(n)}_l r^{(n)}_k   |\< \phi^{(n)}_l|
\psi^{(n)}_k\>|^2.
\end{eqnarray}
The operator $A_n$ has degenerated eigenvalues, which makes its
set of eigenvectors not unique, i.e., any linear combination of
two eigenvectors associated with the same eigenvalue is an
eigenvector itself. Thus, finding the basis
$\lbrace\psi^{(n)}_{k}\rbrace$, in which both $A_n$ and
$(\rho^{\otimes  n})'$ are diagonal, seems to be a difficult
problem that depends on the particular form of $\rho$. So this
approach is not universal. However, now we can express $A_n=P_n^+
- P_n^-$ in terms of the projectors onto the symmetric ($P_n^+$)
and antisymmetric ($P_n^-$) subspaces. From the above it follows
that
\begin{equation}
\Pi_n = \tr[B_n P_n^+(\rho^{(n)})' P_n^+] - \tr[B_n
P_n^-(\rho^{(n)})' P_n^-].
\end{equation}
Moreover we can see that
\begin{equation}
P_n^\pm(\rho^{(n)})' P_n^\pm = P_n^\pm\rho^{(n)} P_n^\pm.
\end{equation}
Thus, we derive
\begin{equation}
\Pi_n = \tr[B_n P_n^+\rho^{(n)} P_n^+] - \tr[B_n P_n^-\rho^{(n)}
P_n^-],
\end{equation}
which corresponds to Eq.~(\ref{eq6}).



\begin{thebibliography}{10}
\bibitem{Schr} E. Schr\"odinger, Proc. Camb. Philos. Soc. {\bf 31}, 555 (1935).



\bibitem{EPR} A. Einstein, N. Podolsky, and B. Rosen, Phys. Rev. {\bf 47}, 777 (1935).



\bibitem{BZ}
I. Bengtsson and K. \.Zyczkowski, \emph{Geometry of Quantum
States} (Cambridge University Press, Cambridge, 2006).



\bibitem{RMP} R. Horodecki, P. Horodecki, M. Horodecki, and K. Horodecki,
Rev. Mod. Phys. {\bf 81}, 865 (2009).



\bibitem{GT} O. G\"uhne and G. T\'oth, Phys. Rep. {\bf 474}, 1 (2009).



\bibitem{PHEkert}
P. Horodecki and A. Ekert, \prl {\bf 89}, 127902 (2002).



\bibitem{PH}
P. Horodecki, \prl {\bf 90}, 167901 (2003).



\bibitem{Carteret}
H. A. Carteret, \prl {\bf 94}, 040502 (2005); quant-ph/0309212.



\bibitem{Horodecki03}
P. Horodecki, \pra {\bf 68}, 052101 (2003).



\bibitem{Bovino}
F. A. Bovino, G. Castagnoli, A. Ekert, P. Horodecki, C.M.
Alves, and A. V. Sergienko, \prl {\bf 95}, 240407 (2005).



\bibitem{Mintert}
A.R.R. Carvalho, F. Mintert, and A. Buchleitner \prl {\bf 93}, 230501 (2004); F. Mintert, M. Ku\'s, and  A. Buchleitner, \prl {\bf 95}, 260502 (2005);



\bibitem{ZHChen}
Z.-H. Chen, Z.-H. Ma, O. G\"uhne, and S. Severini, \prl {\bf 109}
200503 (2012).



\bibitem{Aolita}
L. Aolita and F. Mintert, \prl {\bf 97}, 050501 (2006).



\bibitem{Walborn}
S. P. Walborn, P. H. Souto Ribeiro, L. Davidovich, F. Mintert, and
A. Buchleitner, Nature (London) {\bf 440}, 1022 (2006).



\bibitem{Badziag}
P. Badziag, C. Brukner, W. Laskowski, T. Paterek, and M.
\.{Z}ukowski, \prl {\bf 100}, 140403 (2008).



\bibitem{Huber10}
M. Huber, F. Mintert,  A. Gabriel, and B. C. Hiesmayr,
\prl {\bf 104},  210501 (2010).



\bibitem{GRW07} O. G{\"u}hne, M. Reimpell, and R. F. Werner,
\prl {\bf 98},  110502 (2007).



\bibitem{EBA07} J. Eisert, F. Brandao, and K. Audenaert,
New J. Phys. {\bf 9}, 46 (2007).



\bibitem{OH10} A. Osterloh and P. Hyllus,
\pra {\bf 81}, 022307 (2010).



\bibitem{Sheng15} 
L. Zhou and Y.-B. Sheng, \pra {\bf 90}, 024301 (2014).

\bibitem{RHZ11} {\L}. Rudnicki, P. Horodecki, and K. \.{Z}yczkowski,
\prl {\bf 107}, 150502 (2011).



\bibitem{JMG}
B. Jungnitsch, T. Moroder, and O. G\"uhne, \prl {\bf 106}, 190502
(2011).



\bibitem{RPHZ12} {\L}. Rudnicki, Z. Pucha\l{}a, P. Horodecki, and K.
\.{Z}ycz\-kowski, \pra {\bf 86}, 062329 (2012).



\bibitem{RPHZ14}
{\L}. Rudnicki, Z. Pucha\l{}a, P. Horodecki, and K.
\.{Z}ycz\-kow\-ski, J. Phys. A: Math. Theor. {\bf 47}, 424035
(2014).


\bibitem{Park}
H. S. Park, S. S. B. Lee, H. Kim, S. K. Choi, and H. S. Sim, \prl
{\bf 105}, 230404 (2010).



\bibitem{Laskowski}
W. Laskowski, D. Richart, C. Schwemmer, T. Paterek, and H.
Weinfurter, \prl {\bf 108} 240501 (2012).




\bibitem{Augusiak08}
R. Augusiak, M. Demianowicz, and P. Horodecki, \pra {\bf 77},
030301 (2008).



\bibitem{Wootters98}
W. K. Wootters, \prl {\bf 80}, 2245 (1998).



\bibitem{Bartkiewicz13fidelity}
K. Bartkiewicz, K. Lemr, and A. Miranowicz, \pra {\bf 88}, 052104
(2013).



\bibitem{Bartkiewicz13discord}
K. Bartkiewicz, K. Lemr, A. \v{C}ernoch, and J. Soubusta, \pra
{\bf 87}, 062102 (2013).

\bibitem{Makhlin00}
Y. Makhlin,  Quantum Inf. Process. \textbf{1},  243 (2002).

\bibitem{Bartkiewicz15}
K. Bartkiewicz, J. Beran, K. Lemr, M. Norek, and A. Miranowicz,
\pra {\bf 91}, 022323 (2015).



\bibitem{Barrett05}
S.~D. Barrett and P. Kok, \pra {\bf 71}, 060310 (2005).



\bibitem{Bartkiewicz13chsh}
K. Bartkiewicz, B. Horst, K. Lemr and A. Miranowicz, \pra {\bf
88}, 052105 (2013).



\bibitem{Marcikic02}
I. Marcikic, H. de Riedmatten, W. Tittel, V. Scarani, H.
Zbinden, and N. Gisin, \pra {\bf 66}, 062308 (2002).



\bibitem{Pittman05}
T.~B.~Pittman, B.~C.~Jacobs, and J.~D.~Franson, \pra~\textbf{71},
032307 (2005).



\bibitem{Kiesel05}
N.~Kiesel, C.~Schmid, U.~Weber, R.~Ursin, and H.~Weinfurter,
\prl~\textbf{95}, 210505 (2005).



\bibitem{Bula13QND}
M. Bula, K. Bartkiewicz, A. \v{C}ernoch, and K. Lemr,
\pra{\textbf{87}}, 033826 (2013).



\bibitem{Eisert00}
J. Eisert, K. Jacobs, P. Papadopoulos, and M. B. Plenio,
\pra~\textbf{62}, 052317 (2000).



\bibitem{Pittman03}
T. B. Pittman, M. J. Fitch, B. C. Jacobs, and J. D. Franson,
\pra~\textbf{68}, 032316 (2003).



\bibitem{Lemr_PRL106}
K.~Lemr, A.~\v{C}ernoch, J.~Soubusta, K.~Kieling, J.~Eisert, and
M.~Du\v{s}ek, \prl~\textbf{106}, 013602 (2011).



\bibitem{Kok02}
P.~Kok, H.~Lee, and J.~P.~Dowling, \pra~\textbf{66}, 063814
(2002).

\end{thebibliography}
\end{document}